\documentstyle[preprint,prl,aps,epsfig]{revtex}
\tighten
\begin{document}
\draft
\title{Comparison of Dirac and microscopic Schr\"odinger optical model
potentials for nucleon-nucleus scattering}
\author{S. Karataglidis and D. G. Madland}
\address{Theoretical Division, Los Alamos National Laboratory, Los
Alamos, New Mexico, 87545}
\preprint{LA-UR-01-1496}
\date{\today}
\maketitle
\begin{abstract}
Detailed comparisons are made between results of calculations for
intermediate energy nucleon--$^{208}$Pb scattering using optical
potentials obtained from Dirac phenomenology and a microscopic
Schr\"odinger model. The two approaches yield quite similar results
for all of the observables with each projectile, including spin
observables, suggesting the two models are physically equivalent. 
A new phenomenon in the analyzing power is confirmed in the results 
for 100~MeV scattering.
\end{abstract}
 
The physics of the interaction of a nucleon with the nucleus has
traditionally been represented by the optical potential. Once the
optical potential is specified, and thus the scattering matrix, all
observables may be calculated. For intermediate energies there have
been a number of methods of calculating the optical potential, the
most successful of which have been the Dirac phenomenology \cite{Ra92}
and a recent microscopic, coordinate space, Schr\"odinger model
\cite{Review}. Both approaches have had success in not only predicting
differential cross sections, but spin and integral observables as well
\cite{Review,Ko89,Ko93,De01}. Note that these models are fundamentally
different. The phenomenological Dirac model is dependent on the
fitting of data to determine the parameters in the assumed
potentials. The microscopic Schr\"odinger model is derived from the
bare nucleon-nucleon ($NN$) potential to obtain an effective
interaction in-medium that when folded with a suitable representation
of the target density gives the optical potential. While a
Schr\"odinger-like potential may be derived from the phenomenological Dirac
one, there is no {\em a priori} reason to assume that that
nonrelativistic potential resembles in any way the potential obtained
from the microscopic model.
 
Yet no real comparison has been made of the results from the Dirac and
microscopic Schr\"odinger approaches for both projectiles
simultaneously.  An earlier comparison of the Dirac and
phenomenological Schr\"odinger models showed problems inherent in the
latter \cite{Ko89}.  The phenomenological Schr\"odinger model usually
assumes a local Woods-Saxon form of the optical potential, whose
parameters are determined by fitting a complete set of data. Such an
approach generally fails to reproduce the spin observables due to the
assumptions made in defining the spin-orbit potential and in the
neglect of nonlocality in the potential, arising microscopically from
the antisymmetrization of the wave function of the projectile and
bound state nucleon in the initial state. The purpose of this Letter
is to compare in detail the Dirac phenomenology and microscopic
Schr\"odinger models for both proton and neutron scattering in the
energy region in which they have had success. The case we have chosen
is that of nucleon elastic scattering from $^{208}$Pb at
100--300~MeV. Previous work has been done on this system at this
energy range in the Dirac approach \cite{Ko90}, from which a striking
phenomenon was identified in the analyzing power at 100~MeV: the
proton and neutron analyzing powers are exactly out of phase with each
other and the one is nearly the inverse of the other about their mean.
 
The Dirac phenomenology begins with spherically symmetric complex
Lorentz isoscalar-scalar and isoscalar-vector potentials together with
the Coulomb potential leading to a Walecka-like scattering model
\cite{SW86} with a large attractive scalar potential and an almost as
large repulsive vector potential.  A second-order reduction of the
Dirac equation then leads to a Schr\"odinger-equivalent equation with
physically correct effective central and spin-orbit potentials by
which the observables are accurately predicted. The spin-orbit term
and a Coulomb correction term (accounting for part of the difference
between proton and neutron in-medium projectile motion) appear
naturally.  Projectile and target isospin dependences are treated by
introducing corresponding spherically symmetric complex Lorentz
isovector-scalar and isovector-vector potentials in a relativistic
generalization of the standard Lane model \cite{Cl84}.  The resulting
Dirac equation is suitable for simultaneous analyses of proton-nucleus
and neutron-nucleus scattering data up to several GeV, provided that
the parameters of the potential have been determined by least-squares
adjustment to existing experimental data \cite{Sc82,Hu88}. Such was
done by Kozack and Madland \cite{Ko90} for nucleon--$^{208}$Pb
scattering, using energy-independent symmetrized Woods-Saxon form
factors, and those results are reproduced herein.
 
The microscopic Schr\"odinger approach \cite{Review} does not assume
any phenomenological form for the potential nor is it derived from any
phenomenological potential. It begins instead with the $g$ matrices of
the $NN$ potential; those $g$ matrices are solutions of the
Brueckner-Bethe-Goldstone equation in infinite nuclear matter. The
Bonn-B $NN$ potential \cite{Ma87} was chosen as the starting point for
the calculations presented herein. A local density approximation is
used to map the infinite matter solution to the nucleus in question by
which an effective $NN$ interaction is defined in medium. When folded
with an appropriate ground state density of the target, the
microscopic optical potential is obtained naturally, incorporating
Pauli blocking and density dependences. The (coordinate space)
potential contains both direct and exchange parts, with the exchange
terms arising from the antisymmetrization of the projectile and bound
state nucleon wave functions, and so the potential is fully
nonlocal. There are no parameters in the model which must be adjusted
after the fact to achieve a reasonable fit: all results are obtained
from a single predictive calculation. We use the code DWBA98
\cite{Ra99} to calculate the optical potential and
observables. Success has been achieved predicting the observables from
proton-nucleus scattering for a number of nuclei across a range of
energies \cite{Review}. For the present case, the ground state density
for $^{208}$Pb was obtained from a Skyrme-Hartree-Fock calculation by
Brown \cite{Br00}. As the $g$ matrix is defined for all two-body spin
and isospin channels, the isospin of the projectile selects the
correct components of the matrix to define the appropriate optical
potential for that projectile. As such the same $g$ matrix defines the
optical potentials for both proton and neutron scattering in a natural
way. The Coulomb interaction is, of course, also included in the
calculations of proton scattering.
 
\begin{figure}
\centering\epsfig{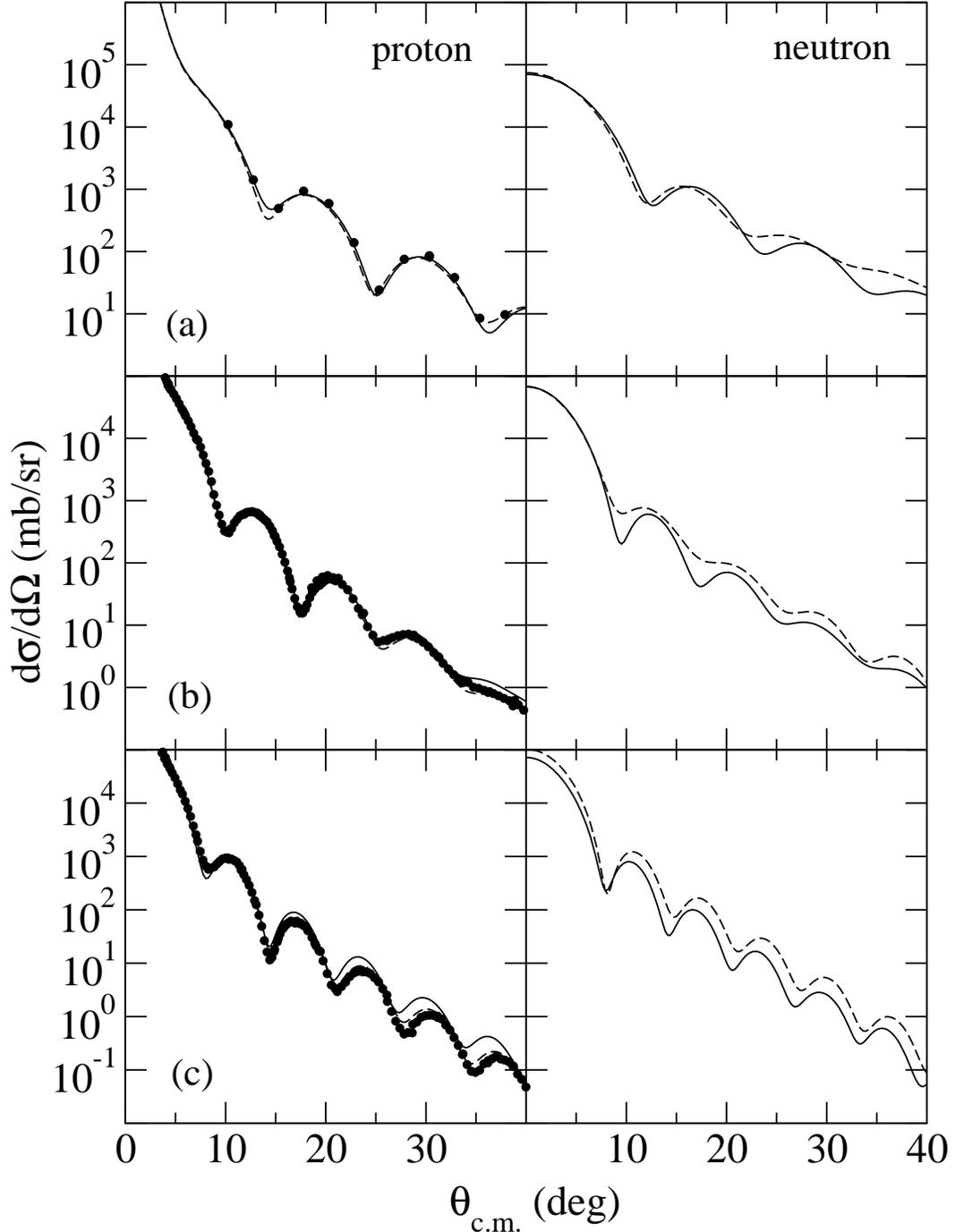}
\caption[]{Differential cross sections for nucleon--$^{208}$Pb elastic
scattering at 100 (a), 200 (b) and 300~MeV (c). The results for proton
and neutron scattering calculated from the Schr\"odinger model are
shown by the solid lines while the results from the Dirac model are
shown by the dashed lines.  The 100~MeV results are compared to the
98~MeV proton scattering data of Schwandt {\em et al.}
\cite{Sc82}. The 200 and 300~MeV proton scattering data are those of
Hutcheon {\em et al.} \cite{Hu88}.}
\label{xsecs}
\end{figure}

In Fig.~\ref{xsecs}, we compare the differential cross sections for
100 [\ref{xsecs}(a)], 200 [\ref{xsecs}(b)] and 300~MeV
[\ref{xsecs}(c)] proton and neutron scattering from $^{208}$Pb from
the Dirac and microscopic Schr\"odinger models. (Hereafter, for
simplicity, we will use the term Schr\"odinger model as referring to
the microscopic Schr\"odinger model.) The results for proton and
neutron scattering calculated from the Schr\"odinger model are shown
by the solid lines, while those from the Dirac model are portrayed by
the dashed lines.  We first consider the results for 100~MeV
scattering. The cross sections for proton scattering as obtained from
both models are very similar in shape and magnitude, and agree quite
well with the 98~MeV data of Schwandt {\em et al.} \cite{Sc82}. While
the similarity between the two models is also observed for neutron
scattering, the level of agreement worsens above $20^{\circ}$,
although both models predict a lack of structure in this angular
region compared to the proton scattering results. The results of the
Schr\"odinger model calculations reproduce the 200~MeV data of
Hutcheon {\em et al.} \cite{Hu88} well, but not so well for the
300~MeV case. This energy is currently the upper limit for the
nonrelativistic microscopic formalism using a bare $NN$ potential. The
underlying $NN$ potential must account for particle production at
300~MeV and higher as each nucleon resonance is excited before
agreement in the nucleon-nucleus scattering may be achieved. This may
account for the disagreement in magnitude between the two models in
the neutron scattering at 300~MeV. Over all energies for both
projectiles, however, the agreement in the calculated differential
elastic scattering from the two approaches is surprisingly good.
 
\begin{figure}
\centering\epsfig{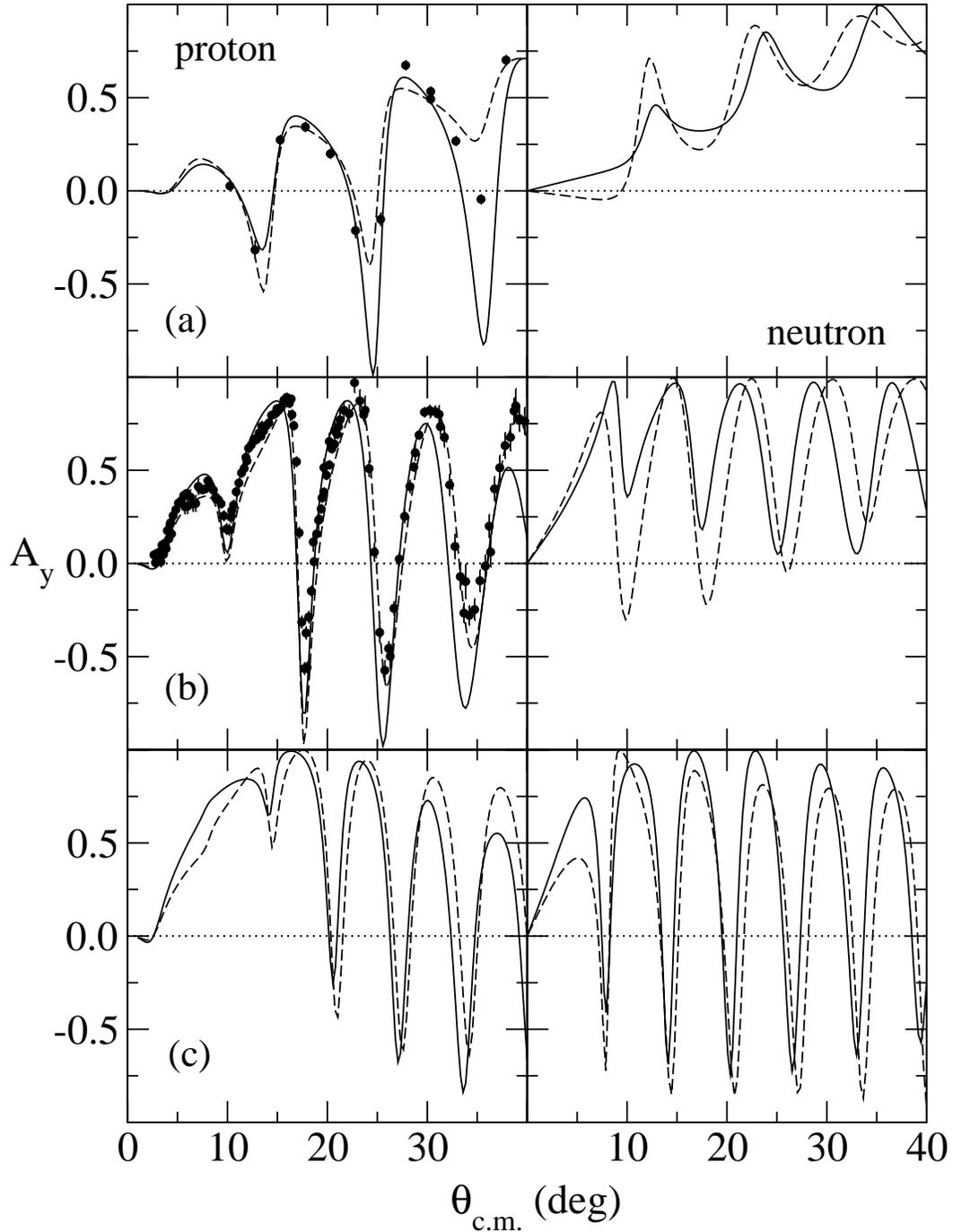}
\caption[]{As for Fig.~\ref{xsecs}, but for the analyzing powers. The
100~MeV results are compared to the 98~MeV proton scattering data of
Schwandt {\em et al.} \cite{Sc82}. The 200~MeV proton scattering data
are those of Hutcheon {\em et al.}  \cite{Hu88}.}
\label{ays}
\end{figure}

\begin{figure}
\centering\epsfig{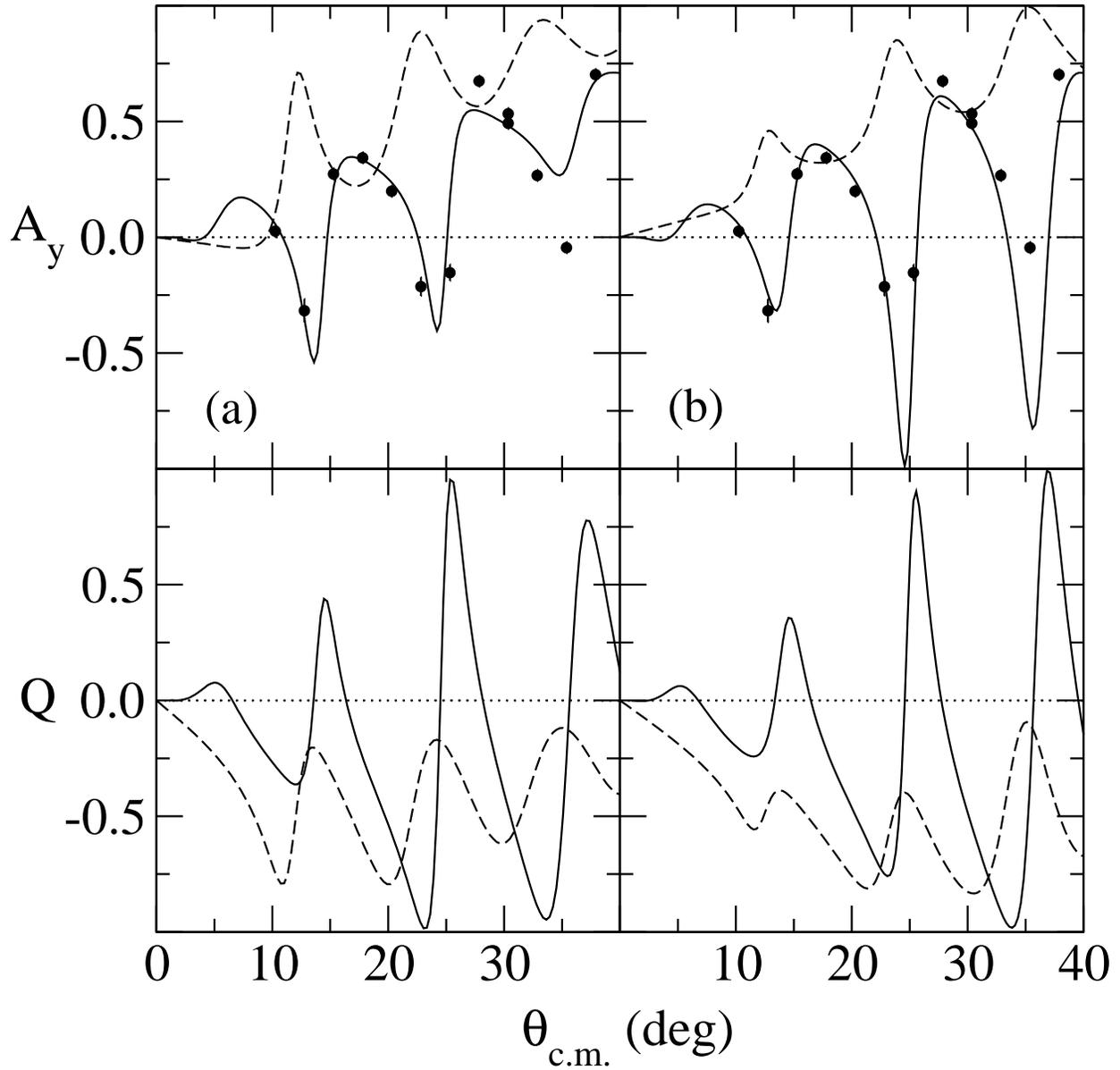}
\caption[]{Analyzing power and spin rotation for the scattering of
100~MeV nucleons from $^{208}$Pb. The Dirac and Schr\"odinger results
are displayed in (a) and (b), respectively, while the proton and
neutron scattering results are portrayed by the solid and dashed
lines, respectively. The proton scattering results are compared to the
98~MeV data of Schwandt {\em et al.} \cite{Sc82}.}
\label{lot}
\end{figure}

As already stated, Kozack and Madland \cite{Ko90} observed a striking
phenomenon in the analyzing powers for 100~MeV nucleon--$^{208}$Pb
elastic scattering. In that work, they noticed that the proton and
neutron scattering analyzing powers were out of phase, and the one
almost the inverse of the other about their mean, with the neutrons
exhibiting a high polarization above $20^{\circ}$ (Fig. 3 of
\cite{Ko90}).  (A detailed investigation of this phenomenon is outside
the scope of this Letter, but work in progress indicates the dominant
element to be the presence or absence of a strong Coulomb field.)  We
compare the results for the analyzing powers at 100, 200, and 300~MeV
from the Schr\"odinger and Dirac models in Fig.~\ref{ays}. In
Fig.~\ref{ays}(a), one observes immediately that the same phenomenon
as observed by Kozack and Madland is reproduced by the Schr\"odinger
model, although the peaks are more exaggerated than in the Dirac
results. Note that both models reproduce the 98~MeV data of Schwandt
{\em et al.} \cite{Sc82}. The phenomenon is explicitly illustrated in
Fig.~\ref{lot}, which displays the spin observables for 100~MeV
scattering. Both the Dirac and Schr\"odinger models predict the
effect. A significant difference between the 100~MeV proton and
neutron scattering spin rotations is also observed with both models,
though not as dramatic as that between the analyzing powers. Their
similarity again gives confidence in the results obtained from both
models and we encourage measurements of these spin observables,
especially the analyzing powers, to see if this is a real effect. Such
a measurement would also serve to assess the models, especially for
neutron scattering. We note here that neither model has yet explicitly
included Mott-Schwinger scattering, but that would be the next step
especially if contrary experimental evidence appears.  The 200 and
300~MeV results are less striking in the proton and neutron analyzing
power differences and beyond $15^{\circ}$ the proton and neutron
analyzing powers are similar for both energies. Significant
differences are observed only in the forward angle scattering,
especially at 300 MeV, where in both models the minimum at $8^{\circ}$
in the neutron scattering analyzing power is missing in that for
proton scattering. Again, the agreement between the two models,
especially in this spin observable, is much better than had been
expected.  Note that the 200~MeV proton scattering data were used in
the global least-squares adjustment to determine the Dirac optical
potential at that energy, and so the level of agreement between those
data and the Dirac result is to be expected. The Schr\"odinger 200~MeV
proton result is a prediction.
 
\begin{figure}
\centering\epsfig{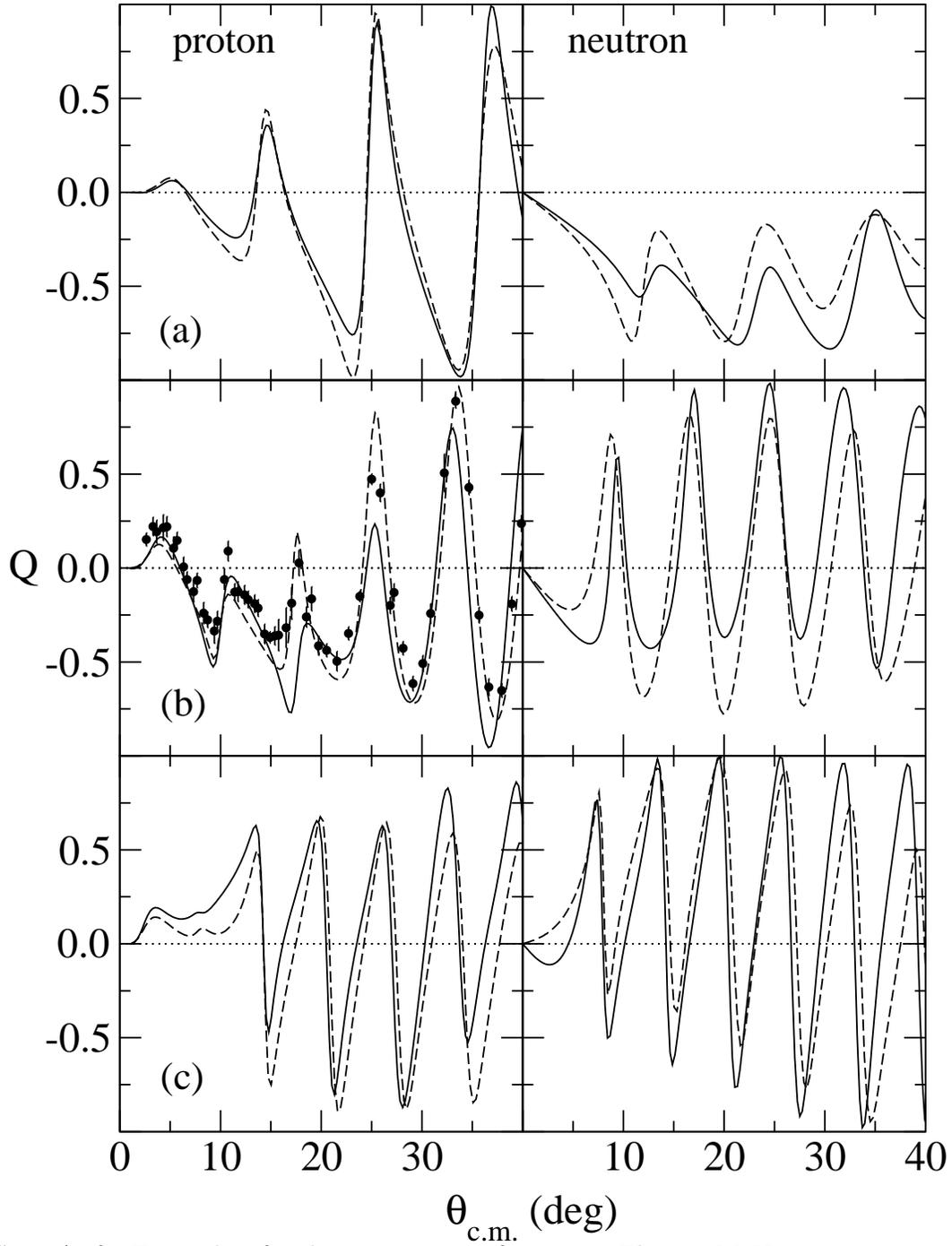}
\caption[]{As for Fig.~\ref{xsecs}, but for the spin rotation
functions. The 200~MeV proton scattering data are those of Hutcheon
{\em et al.}  \cite{Hu88}.}
\label{spinq}
\end{figure}

The dramatic difference in the analyzing power at 100~MeV is also seen
in the spin rotation function, displayed in Fig.~\ref{spinq}. As with
the other two observables, the relative features between the proton
and neutron scattering results at all three energies are common to
both models. However, the variations in the neutron scattering results
from the two models are more distinct at 100~MeV, and less so as one
increases the energy. At 200~MeV, the agreement between the Dirac
model and the data of Hutcheon {\em et al.} \cite{Hu88} again is quite
good. While that level of agreement is not obtained with the
Schr\"odinger model that calculation reproduces most of the features
in the data. As with the analyzing powers at 200 and 300~MeV, the
differences between the proton and neutron spin rotation functions
from both models significantly differ only below $20^{\circ}$.  Thus,
the two model approaches are in good agreement for both spin
observables for both projectiles in their similarities as well as in
their differences.
 
Fundamentally, our Dirac and Schr\"odinger approaches are
different. That the two models agree so well for the differential
cross sections as well as both spin observables for both projectiles
not only gives confidence in the calculated results at the three
energies considered but also suggests, rather strongly, that the
models are physically equivalent over this energy range. Namely, the
microscopic Schr\"odinger model incorporates all the dominant medium
modifications in the optical potential without significant
approximation by using a realistic ground state density yielding a
reasonable specification of all terms in the optical potential,
whereas the Dirac approach provides for a natural specification of
such terms. The comparison obtained in this work is exemplary of the
dilemma in judging the relative merits of relativistic {\em vs.}
nonrelativistic approaches in intermediate energy proton-nucleus
scattering analyses \cite{Ra92}. As Ref. \cite{Ra92} speculates, the
answer throughout the energy regime may lie in QCD-based models of
nuclear scattering systems. Such theoretical models and concomitant
experiments will require simultaneous treatment of proton and neutron
scattering in order to be complete; the latter's experimental database
is currently sadly lacking. Finally, the question remains: is the
difference between the proton and neutron analyzing powers at 100~MeV
real? Only experiment will provide the final answer to that question.

This work was supported by the United States Department of Energy
Contract no. W-7405-ENG-36. It is dedicated to the memory of
Richard Kozack.

%
%
 
\end{document}